\documentclass{aa}

\usepackage{graphicx}
\usepackage{caption}
\usepackage{subcaption}
\usepackage{multirow}
\usepackage{txfonts}
\usepackage{xcolor}
\usepackage[breaklinks,colorlinks,urlcolor=blue,citecolor=blue,linkcolor=blue]{hyperref}
\usepackage{natbib}
\defcitealias{2026A&A...705A.157O}{Paper~I}

\begin{document}

   \title{Characterization of white-light enhancements under umbral conditions in one-dimensional simulations of solar flares}

   \author{S. Ornig\inst{1} \fnmsep \inst{2}\and
          M. Carlsson\inst{1} \fnmsep \inst{2}}

   \institute{Institute of Theoretical Astrophysics, University of Oslo,
              PO Box 1029 Blindern, 0315 Oslo, Norway\\
              \email{sascha.ornig@astro.uio.no}
         \and
             Rosseland Centre for Solar Physics, University of Oslo,
              PO Box 1029 Blindern, 0315 Oslo, Norway\\
             }

   \date{Received 01 May 2026; accepted 03 June 2026}

  \abstract 
   {Solar flares with signatures in the optical continuum (white light, WL) pose a challenge not only to the standard flare model, but also to solar flare simulations. In particular, simulations are so far not able to convincingly reproduce observed WL enhancements.}
   {We investigate the effect of different electron beams on an umbral atmosphere and what the differences and similarities to the quiet-Sun response are.}
   {We characterized WL emission in one-dimensional simulations of solar flares using the radiation hydrodynamics code RADYN. We created such simulations with a similar setup as the \mbox{F-CHROMA} grid, but with a starting atmosphere describing umbral conditions. We investigated the influence of different temporal profiles of an electron beam on this umbral atmosphere.}
   {Our simulations show maximum WL increases between 40 and 335\%, which is comparable to observed values. The reduced umbral background is the main reason for these large increases. We identify hydrogen recombination in an optically thin chromosphere as the dominant process responsible for the increases, with the radiation from the heated photosphere becoming substantial in the later stages (a few seconds after beam heating is turned off) due to the longer timescale of the cooling of the photosphere compared to hydrogen recombination in the chromosphere. The short timescale of recombination results in very short gradual phases of excess emissions in our simulations. Shorter, more intense beams (i.e., beams with a higher maximum energy flux) lead to a faster and more dramatic atmospheric evolution. Such beams also cause larger WL enhancements due to a higher electron density in the relevant layers. Both the Balmer ratio and the Paschen ratio are substantially higher in our simulations compared to simulations with a quiet-Sun atmosphere.}
   {The detectability and amplitude of excess WL emissions depends both on the spectral and temporal structure of the electron beam impacting on the atmosphere as well as the underlying background radiation. The combination of a short, intense beam and an umbral atmospheric structure provides an excellent seed for substantial WL enhancements.}

   \titlerunning{White-light enhancements under umbral conditions in flare simulations}
   \maketitle


\section{Introduction}

Energy release processes in the solar atmosphere are of high importance for the understanding of the Sun in general. One example of transient energetic phenomena on the Sun is a solar flare, which manifests itself as a short-lived, localized brightening. In the standard flare model \citep{1964NASSP..50..451C,1966Natur.211..695S,1974SoPh...34..323H,1976SoPh...50...85K}, solar flares are thought to be caused by magnetic reconnection, resulting in, among other processes, acceleration of charged particles (in the form of a beam) toward the lower atmospheric layers. Here, the beam particles deposit their energy, heating and ionizing the ambient plasma. The increase in gas pressure leads to expansion of the plasma \citep[termed chromospheric evaporation, see][]{1985ApJ...289..414F}, which subsequently fills up the reconnected magnetic loop. The evaporated plasma then cools down by radiating energy in the soft X-ray (SXR) and extreme ultraviolet (EUV) spectral ranges \citep{2017LRSP...14....2B}.\\
\indent Spectral lines provide a plethora of insights into the mechanisms behind solar flares, and the majority of studies focus on such lines in the electromagnetic spectrum. Nonetheless, the earliest solar flare observations were performed in the optical continuum \citep{1859MNRAS..20...13C,1859MNRAS..20...15H}, known as white light (WL). Accordingly, white-light flares (WLFs) are solar flares that show signatures in this wavelength range. The term ``optical'' in association with WL is a rather loose term and often also includes parts of the ultraviolet and infrared as well.\\
\indent Based on the proposed mechanisms behind them, WLFs exhibit different temporal and spectral behavior. Type~I WLFs show continuum signatures that are temporally coincident with hard X-ray (HXR) and microwave increases. They also exhibit strong Balmer lines as well as a jump in the continuum intensity: the Balmer jump \citep{1986A&A...159...33M,1994ApJ...429..890D}. Type~I~WLFs are thought to stem from optically thin hydrogen recombination in the chromosphere \citep{2018ApJ...867L..24D}. Type~II~WLFs, on the other hand, lack a jump in the hydrogen continua, have weak and narrow Balmer lines, and show a time discrepancy between HXR and microwave increases and WL enhancements \citep{1995A&AS..110...99F,1999ApJ...512..454D}. The proposed mechanism is enhanced H$^-$ continuum radiation from the photosphere \citep{1989SoPh..121..261N,2014ApJ...783...98K,2016ApJ...816...88K} as a result of photospheric heating either through high-energy charged-particle beams \citep{1978ApJ...224..241E,2018ApJ...862...76P}, Alfvén-wave heating \citep{2013ApJ...765...81R}, or radiative backwarming \citep{1989SoPh..124..303M,2003ApJ...595..483M}.\\
\indent The magnitude of the WL enhancement depends not only on the energy of the flare, but also on the photospheric background. It has been shown in observations that WL increases are larger in the (pen-)umbral region of sunspots than in the quiet Sun \citep{1999ApJ...512..454D,2020ApJ...904...96C,2021RAA....21....1Y,2024SoPh..299...11J}, since the lower temperature in these areas results in lower photospheric background radiation. The enhancements are then more easily detectable and larger (relative to the pre-flare intensity level). Using the radiation hydrodynamics code RADYN \citep{1992ApJ...397L..59C, 1995ApJ...440L..29C, 1997ApJ...481..500C, 2002ApJ...572..626C,2015ApJ...809..104A}, simulations of WL enhancements in penumbral conditions \citep[utilizing the semiempirical penumbral atmosphere of][]{1989A&A...225..204D} were conducted by \cite{2021RAA....21....1Y} and \cite{2023ApJ...952L...6S}. \cite{2021RAA....21....1Y} found continuum contrasts at 4250~\AA~enhanced by up to a factor of 10 compared to simulations with a quiet-Sun starting atmosphere, while the simulations of \cite{2023ApJ...952L...6S} showed a difference of a factor of two to four at 6173~\AA.\\
\indent In this paper, we focus on WL enhancements stemming from electron bombardment of the lower solar atmosphere in umbral conditions using the one-dimensional radiation hydrodynamics code RADYN. This work therefore presents an extension to the \mbox{F-CHROMA} grid of solar flare simulations \citep{2023A&A...673A.150C}. Enhancements of WL emissions in the \mbox{F-CHROMA} grid were investigated by \cite{2026A&A...705A.157O}, and we refer to this as \citetalias{2026A&A...705A.157O} from here on. Here, we aim to quantify the WL enhancements associated with an underlying umbral atmosphere and compare them with the increases seen in simulations with a VAL-C-like (quiet-Sun) atmosphere \citep{1981ApJS...45..635V} as in \citetalias{2026A&A...705A.157O}.
\begin{figure}
    \includegraphics[width=0.48\textwidth]{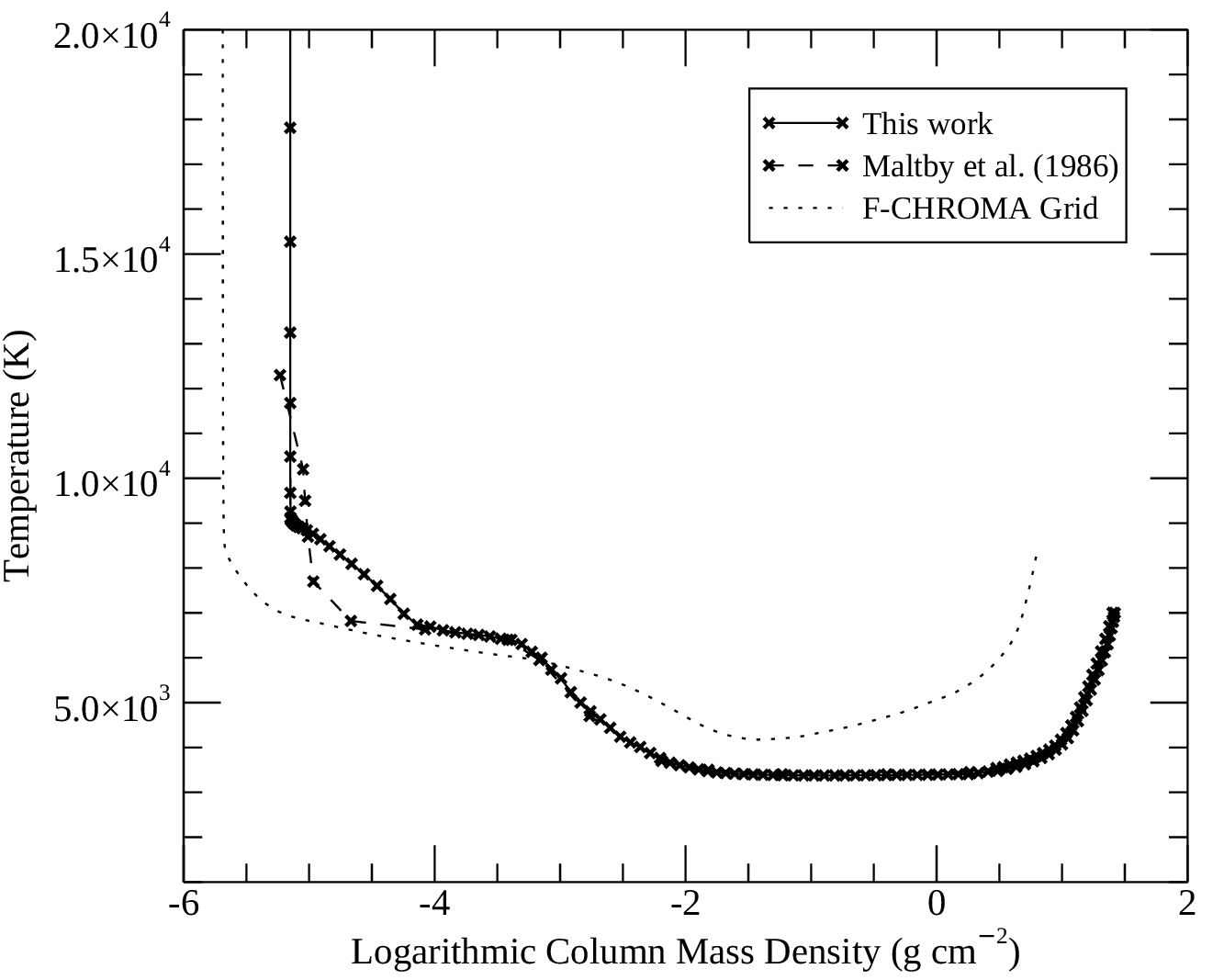}
    \caption{Temperature as a function of column mass density in the RADYN starting atmosphere used in this work (solid line), in Model M of \cite{1986ApJ...306..284M} (dashed line), and in the simulations included in the F-CHROMA grid (dotted line).}
    \label{fig:atm_start}
\end{figure}


\section{Methodology}

\subsection{The radiation hydrodynamics code RADYN}

One of the most widely used codes for simulating the atmospheric response to solar flares is the radiation hydrodynamics code \mbox{RADYN} \citep{1992ApJ...397L..59C, 1995ApJ...440L..29C, 1997ApJ...481..500C, 2002ApJ...572..626C,2015ApJ...809..104A}. It combines the equations of hydrodynamics with the radiative transfer equation and solves them using a Newton-Raphson linearization method on an adaptive grid \citep{1987JCoPh..69..175D}. The version of RADYN we used for our purposes is the same as the one employed in the making of the \mbox{F-CHROMA} grid of models \citep{2023A&A...673A.150C}. The hydrogen atom was modeled with five levels plus a continuum, helium with eight levels plus a continuum, and calcium with five levels plus a continuum. Other elements were treated as background continua in local thermodynamic equilibrium \citep{2002ApJ...572..626C}. Our starting atmosphere was modeled as a quarter-circle loop with a half length of 10~Mm, extending just below the photosphere (approximately 90~km below the \mbox{$\tau_{500nm} = 1$} height). Both the boundary at the top and bottom were closed.\\
\indent As in the \mbox{F-CHROMA} grid, we used an electron beam to heat the atmosphere. The beam electrons were assumed to be nonthermal and to follow a power-law energy distribution, which can be described by three parameters: the spectral index, $d$, the low-energy cutoff (or cutoff energy), $E_c$, and the total beam energy, $E_{tot}$. We used the Fokker-Planck description \citep{2015ApJ...809..104A} to model the impact of these beam electrons on the atmosphere. The timestep size of our simulations spans several orders of magnitude, from roughly $10^{-2}$~s to about $10^{-8}$~s, but the chosen temporal resolution of our output is 0.1~s.\\
\indent In order to be able to compare our simulations with those in the \mbox{F-CHROMA} grid, while maintaining a reasonable scope for this work, we focused our investigations on one particular particle-beam setup, which is the same as in case~2 of \citetalias{2026A&A...705A.157O}. We therefore modeled an electron beam with $d = 3$, $E_{c} = 25$~keV, and $E_{tot} = 10^{12}$~erg~cm$^{-2}$.

\subsection{Starting atmosphere}

To simulate umbral conditions, we made use of the semiempirical umbral core model M presented in \cite{1986ApJ...306..284M}. The temperature minimum region (TMR) in this model has a temperature of 3400~K. The model parameters cover a range from $\log m \approx 1.43 $ to $\log m \approx 5.23$, corresponding to a physical height from 122~km below to 2126~km above the $\tau_{500} = 1$ surface. As this model does not include a transition region (TR) or corona, we artificially added a TR and corona in a similar way as described for the \mbox{F-CHROMA} grid \citep{2023A&A...673A.150C}, but with a top temperature of $2\cdot10^6$~K. This temperature is needed in order to better fit the location of the TR (on a column-mass scale) with that of \cite{1986ApJ...306..284M}. We further justify this high temperature via the fact that we are simulating a sunspot in an active region (AR), and AR temperatures are generally higher than those in the quiet Sun.\\
\indent An issue that arises when implementing the model M atmosphere of \cite{1986ApJ...306..284M} is that RADYN uses a loop geometry, meaning that the gravitational acceleration varies along the loop (being zero at the top boundary). The density scale height is therefore different to that in \cite{1986ApJ...306..284M}, and deviations in either the height scale or the column mass scale result. We chose to adopt the same temperature variation on a column mass scale as in \cite{1986ApJ...306..284M}. As Fig.~\ref{fig:atm_start} shows, the fit in the photosphere and lower chromosphere is very good, but the top of the chromosphere experiences too much heating (from the incoming radiation) in our setup. Figure~\ref{fig:atm_start} also contains the starting atmosphere for the F-CHROMA simulations (i.e., quiet Sun) as a reference.

\subsection{Beam profile shapes}

\begin{figure}
    \includegraphics[width=0.48\textwidth]{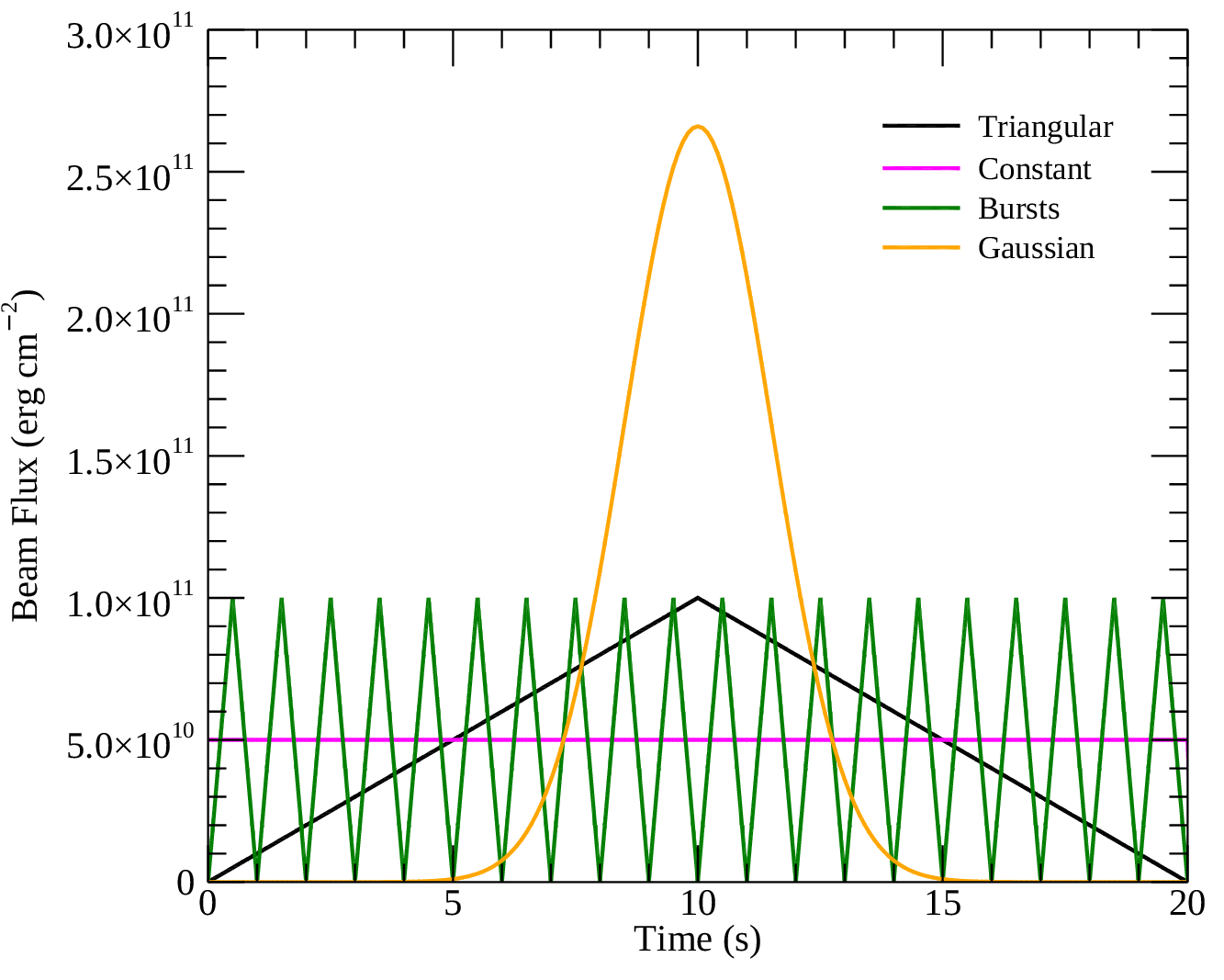}
    \caption{Temporal profiles of the electron beams used in this work. We note that the amplitude and duration differs between simulations (as indicated in Tab.~\ref{tab:sims}).}
    \label{fig:beam_profiles}
\end{figure}

The \mbox{F-CHROMA} grid includes beams with a triangular temporal shape over 20~s, i.e., a linear increase over the first 10~s followed by a linear decrease back to zero over the subsequent 10~s. Here, we want to explore how different temporal profiles of electron beams influence the resulting continuum enhancements. For this, we selected four different beam patterns: (i) a triangular shape similar to the one used in the \mbox{F-CHROMA} grid, (ii) a constant beam flux, (iii) an ensemble of 1-second bursts \citep[to mimic short-period quasi-periodic pulsations, cf.][]{2016SoPh..291.3143V}, and (iv) a Gaussian profile. For illustration, example temporal profiles are shown in Fig.~\ref{fig:beam_profiles}, and the parameters as well as designations used throughout this work can be found in Tab.~\ref{tab:sims}. We ensured that the total beam energy is the same for all simulations ($10^{12}$~erg~cm$^{-2}$).
\newcommand{\specialcell}[2][c]{%
\begin{tabular}[#1]{@{}c@{}}#2\end{tabular}}
\begin{table}
    \caption{Beam parameters and designations of all simulations used in this work.}\label{tab:sims}
    \centering
    \begin{tabular}{ c c c c }
    \hline
    Label & Beam Shape & \specialcell{Max. Flux\\(erg~cm$^{-2}$)} & \specialcell{Beam Duration\\(s)} \\
    \hline
    T20 & Triangular & $10^{11}$ & 20 \\
    T10 & Triangular & $2\cdot10^{11}$ & 10 \\
    C20 & Constant & $5\cdot10^{10}$ & 20 \\
    C10 & Constant & $10^{11}$ & 10 \\
    C5 & Constant & $2\cdot10^{11}$ & 5 \\
    C1 & Constant & $10^{12}$ & 1 \\
    B20 & Bursts & $10^{11}$ & 20 \\
    G20$_{\mathrm{slim}}$ & Gaussian & $2.66\cdot10^{11}$ & 20 \\
    G20$_{\mathrm{wide}}$ & Gaussian & $10^{11}$ & 20 \\
    \hline
    \end{tabular}
    \tablefoot{A beam duration of 20~s for the burst case means 20 one-second triangular bursts.}
\end{table}

\subsection{Relevant (continuum) parameters}

Our chosen continuum wavelength is 6684~\AA, which we selected based on two factors. First and foremost, it is the same wavelength we used in \citetalias{2026A&A...705A.157O}, making comparisons with that work easier. Second, this wavelength is reasonably close to the wavelength used for continuum measurements by the Helioseismic and Magnetic Imager on the Solar Dynamics Observatory \citep{2012SoPh..275..207S}. A comparison of our results using 6684~\AA~and 6175~\AA~(an example of which is shown in the form of the continuum light curves in the appendix) shows that the differences are negligible.\\
\indent Similar to \citetalias{2026A&A...705A.157O}, we made use of the Balmer and Paschen ratios. The ratios were calculated around the respective jump in the continuum, which arises as a result of the ionization energy of the hydrogen atom from different energy states \citepalias[for more details, see][]{2026A&A...705A.157O}. We defined the ratios as
\begin{gather}\label{eq:ratios}
    \hspace{0.2\textwidth} R_i = \frac{I_\mathrm{b,i}}{I_\mathrm{r,i}},
\end{gather}
where the index $i = B,P$ (for Balmer and Paschen), while $I_\mathrm{b}$ is the intensity blueward and $I_\mathrm{r}$ is the intensity redward of the respective jump. We used the wavelengths 3646.9~\AA~(blueward) and 3647.1~\AA~(redward) for the calculation of the Balmer ratio, and 8205.7~\AA~(blueward) and 8206.0~\AA~(redward) for the Paschen ratio.


\section{Results}

\begin{figure}
    \includegraphics[width=0.48\textwidth]{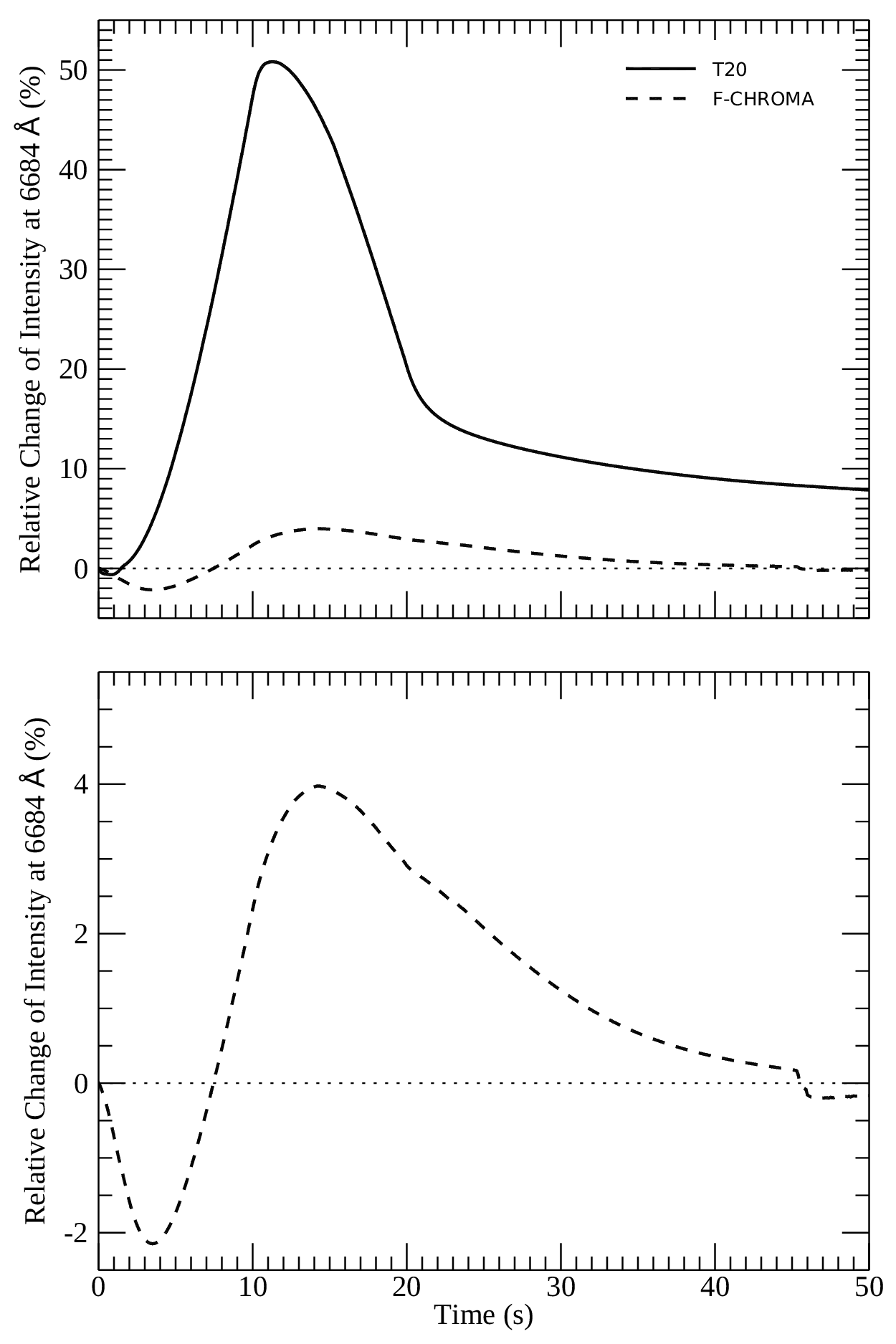}
    \caption{Comparison of the continuum enhancement at 6684~\AA~in our case T20 (solid line) and in the \mbox{F-CHROMA} model with the same setup, but with a quiet-Sun starting atmosphere (dashed line). The bottom panel displays a zoom to better visualize the \mbox{F-CHROMA} case. The horizontal dotted line shows the zero level.}
    \label{fig:compare_umbra_qs}
\end{figure}

\begin{figure}
    \includegraphics[width=0.48\textwidth]{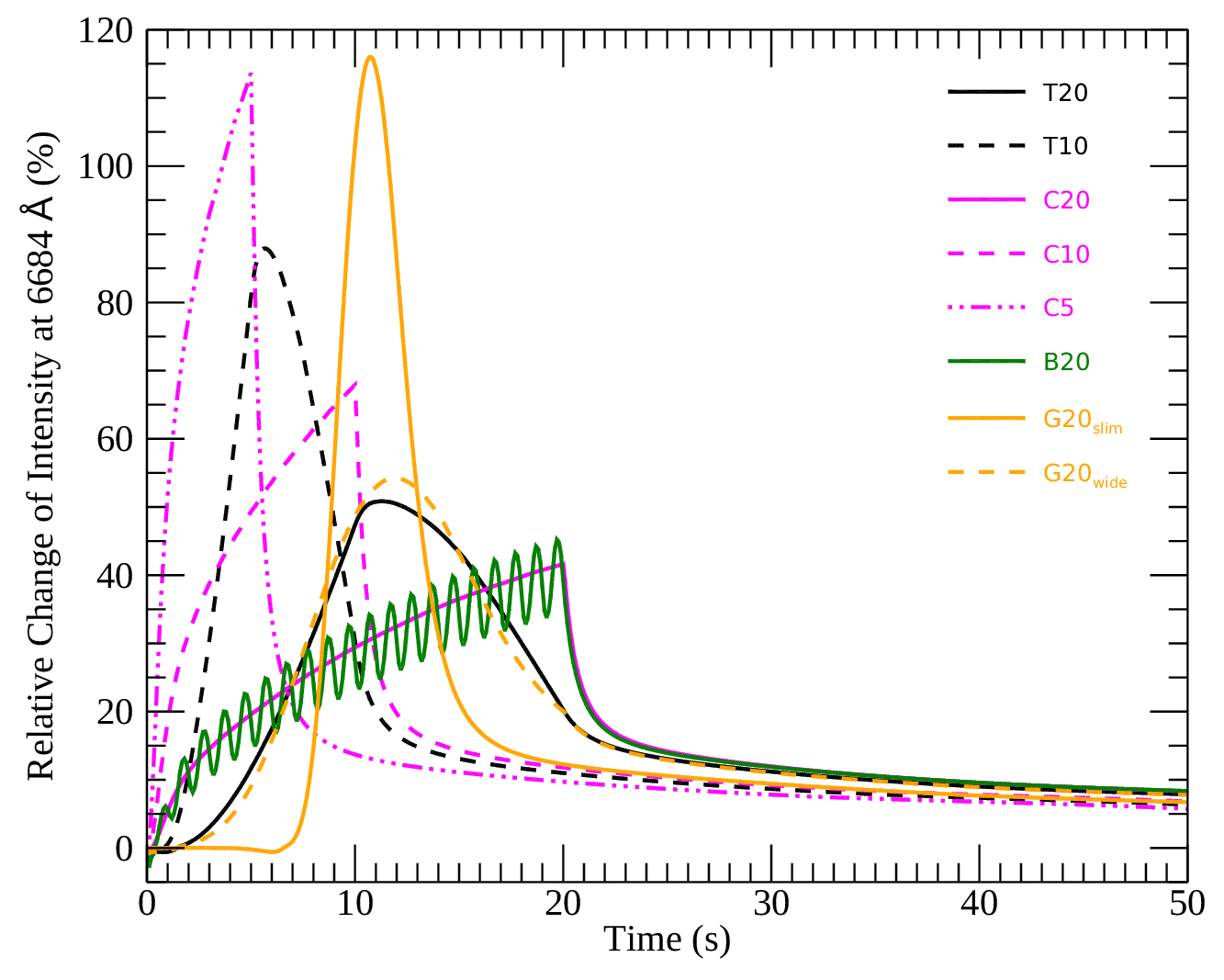}
    \caption{Continuum light curves at 6684~\AA~for all simulations in this work. Intensity increases are given with respect to the intensity at the start of the simulations. Case C1 is not included here in order to keep the y axis range within a desirable level.}
    \label{fig:c6684}
\end{figure}

\begin{figure}
    \includegraphics[width=0.48\textwidth]{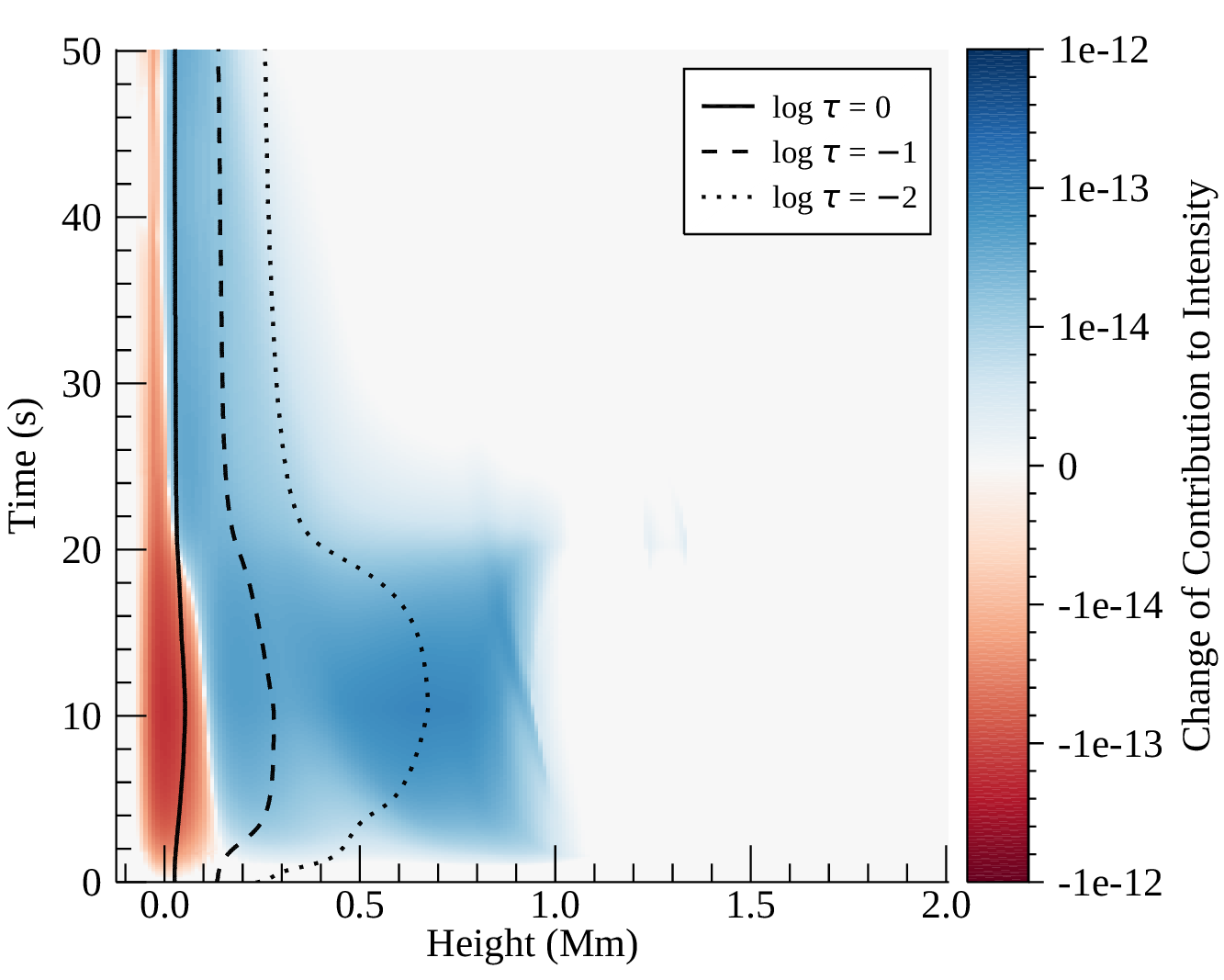}
    \caption{Change of the contribution to the continuum intensity at 6684~\AA~as a function of height and time on a logarithmic scale for case~T20. The scale is displayed as a color bar on the right side. The bin size in the space domain is $10$~km, and $0.1$~s in the time domain. The solid, dashed, and dotted black lines indicate the height where \mbox{$\tau = 1.0$}, $0.1$, and $0.01$, respectively.}
    \label{fig:contrib_tri}
\end{figure}

\begin{figure}
    \includegraphics[width=0.48\textwidth]{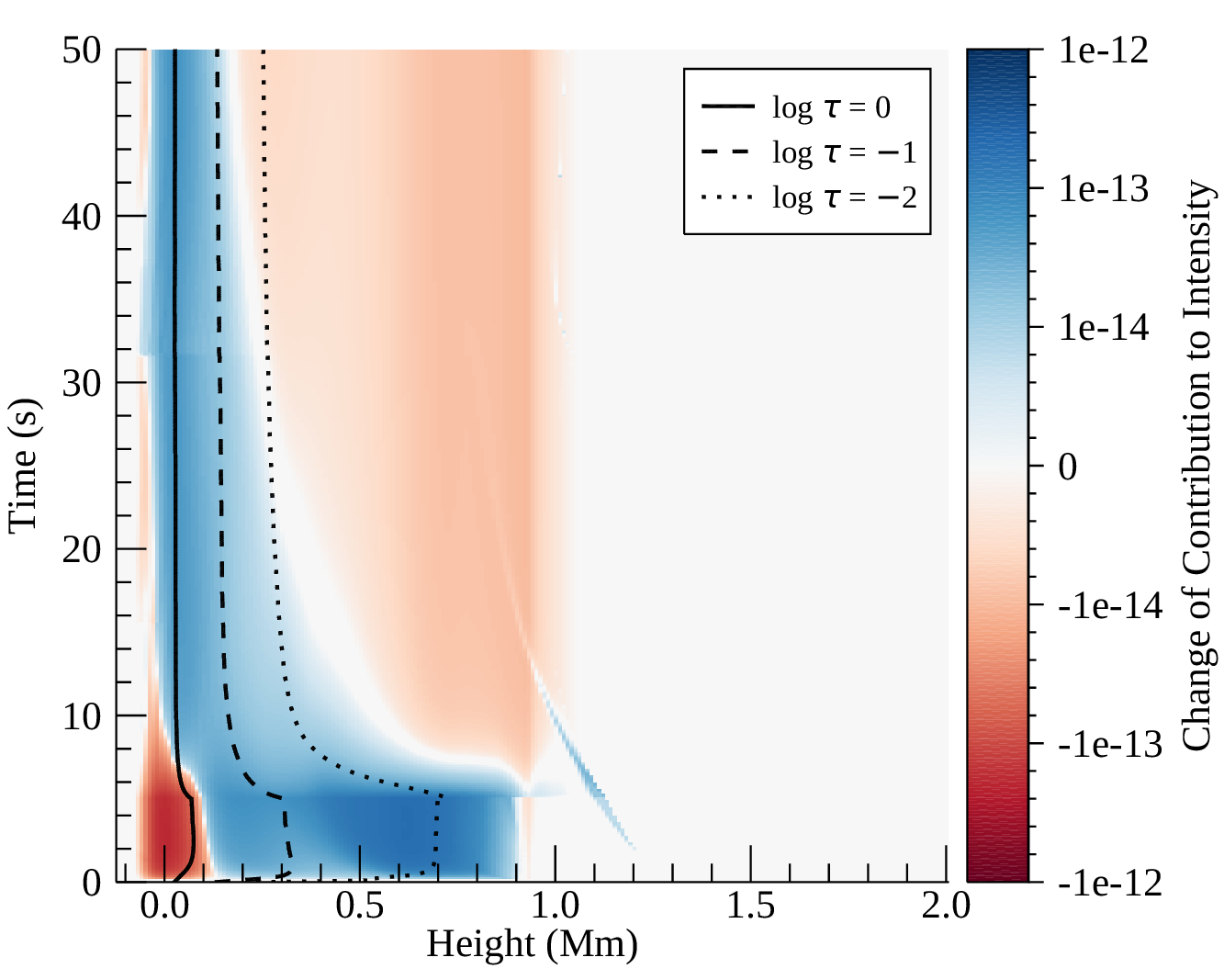}
    \caption{Same as Fig.~\ref{fig:contrib_tri}, but for case C5.}
    \label{fig:contrib_const_5s}
\end{figure}

In general, the atmosphere during the simulated flare-energy input experiences a high level of heating for all our simulations. For example, in case T20 the top of the chromosphere is heated from about 7,000~K to nearly 50,000~K (about 70,000~K in case T10). Even the lower chromosphere reaches almost 8,000~K, up from about 4,000~K at $t = 0$. The TMR is pushed down by over 200~km (or almost one order of magnitude in terms of column mass). Shorter, more intense beams lead to a faster atmospheric evolution. Ionization of the ambient atoms in the chromosphere is faster and slightly more extensive in these cases. As an example, the chromospheric hydrogen in the layers experiencing beam-energy deposition is fully ionized twice as fast in case T10 than in case T20. The largest depth of full ionization lies slightly deeper for T10 than for T20 (by about 50~km). Furthermore, more intense beams cause higher energy levels of hydrogen to be populated more and faster throughout the photosphere and chromosphere, resulting in larger opacities and more efficient stopping of photospheric radiation.\\
\indent The combination of $d$, $E_{cut}$, and $E_{tot}$ under consideration here leads to an increase in the continuum light curve at 6684~\AA~(measured relative to $t = 0$) of 4.0\% with the setup in the \mbox{F-CHROMA} grid \citepalias{2026A&A...705A.157O}, i.e. with a quiet-Sun starting atmosphere and a triangular 20-second beam. Comparing this with our case T20 (Fig.~\ref{fig:compare_umbra_qs}), we notice four aspects. First, the initial decrease in intensity is lower for case T20 (0.6\% vs. 2.1\% in the \mbox{F-CHROMA} case). Second, case T20 reaches its maximum continuum enhancement earlier. Third, the enhancement is an order of magnitude larger (50.8\%). Finally, the decline after the maximum is more modest in the \mbox{F-CHROMA} case. We discuss each of these points in more detail below.\\
\indent The fact that the initial decrease is lower for case T20 than for the \mbox{F-CHROMA} case is due to the change of the hydrogen level population density. In the \mbox{F-CHROMA} case, this change is larger, leading to a higher population density of all excited states of hydrogen. This is despite the larger level population density at $t = 0$ in our case. The larger increase leads to a bigger change of the opacity in the chromosphere, which ends up absorbing more of the photospheric radiation in the \mbox{F-CHROMA} case, leading to a larger decrease in the continuum light curve.\\
\indent All cases under study here show immense continuum increases compared to the simulations included in the \mbox{F-CHROMA} grid. Figure~\ref{fig:c6684} shows the continuum light curves at 6684~\AA~for our cases (except for case~C1, due to its enormous increase compared to the other cases), with the key properties captured in Tab.~\ref{tab:sims_wl}. The triangular 20-second beam leads to an increase of 50.8\%, an order of magnitude larger than in the \mbox{F-CHROMA} grid. Overall, the maximum increases range from 41.6\% (case~C20) to 334.9\% (case~C1). However, we note that the continuum intensity at 6684~\AA~at $t = 0$ plays a big role here, as it is an order of magnitude higher in the \mbox{F-CHROMA} case. Even the largest increases in our simulations (case C1) do not result in intensity levels exceeding the $t = 0$ intensity in the \mbox{F-CHROMA} simulation (the discrepancy is still a factor of two). Comparing the absolute increases, our cases indeed show larger values, but only by up to a factor of ten (for case C1), which is not comparable to the relative increases (which are over 83 times higher for case C1 than for the \mbox{F-CHROMA} case). It follows that the main reason for the much higher relative continuum increases in our simulations compared to the \mbox{F-CHROMA} cases is the much lower background intensity in our umbral atmosphere.\\
\renewcommand{\specialcell}[2][c]{%
\begin{tabular}[#1]{@{}c@{}}#2\end{tabular}}
\begin{table}
    \caption{Properties of the continuum at 6684~\AA~for each simulation.}\label{tab:sims_wl}
    \centering
    \begin{tabular}{ r r r r r r }
    \hline
    Label & \specialcell{$\mathrm{I_{rel,max}}$\\(\%)} & \specialcell{$\mathrm{t_{max}}$\\(s)} & \specialcell{$\mathrm{t_{dec}}$\\(s)} & \specialcell{$\mathrm{C_{phot,max}}$\\(\%)} & \specialcell{$\mathrm{C_{phot,dec}}$\\(\%)}\\
    \hline
    T20 & 50.8 & 11.3 & 9.1 & 7.4 & 31.3 \\
    T10 & 87.9 & 5.7 & 4.2 & 5.4 & 12.5 \\
    C20 & 41.6 & 20.0 & 3.6 & 14.4 & 57.9 \\
    C10 & 67.9 & 10.0 & 1.3 & 8.1 & 24.5 \\
    C5 & 113.6 & 5.0 & 0.8 & 4.8 & 11.7 \\
    C1 & 334.9 & 1.0 & - & 0.4 & - \\
    B20 & 45.2 & 19.7 & 2.7 & 12.7 & 46.3 \\
    G20$_{\mathrm{slim}}$ & 116.0 & 10.7 & 2.7 & 2.1 & 1.8 \\
    G20$_{\mathrm{wide}}$ & 54.2 & 11.9 & 8.2 & 7.7 & 29.3 \\
    \hline
    \end{tabular}
    \tablefoot{$\mathrm{I_{rel,max}}$ describes the maximum intensity increase at 6684~\AA~relative to $t=0$. $\mathrm{t_{max}}$ is the time of the maximum, whereas $\mathrm{t_{dec}}$ is the decay time of enhancements, i.e., the time it takes for the enhancements to drop below $\frac{1}{e}$ of the maximum. $\mathrm{C_{phot,max}}$ and $\mathrm{C_{phot,dec}}$ are the photospheric contribution to the intensity increase at the time of maximum and after the decay time, respectively. The value of $\mathrm{C_{phot,dec}}$ is small for case G20$_{\mathrm{slim}}$ because the $\frac{1}{e}$ level is reached early. Case C1 has not reached past the decay time yet, and we therefore mark the entries relying on the decay time in that row with a minus to signify missing values.}
\end{table}
\indent Some interesting conclusions can be drawn from Fig.~\ref{fig:c6684}. Firstly, we note that there is almost no gradual phase for any of the simulations once beam heating is turned off. This can be seen easiest for the simulations with a constant beam flux. We can describe this via the decay time, i.e., the time it takes the enhancements to drop below the $\frac{1}{e}$ level of the maximum. The values are shown in the fourth column of Tab.~\ref{tab:sims_wl}, and are generally on the order of a few seconds. A similar evolution was found in the infrared by \cite{2017A&A...605A.125S} and \cite{2024MNRAS.532..705S}, although the authors note that the decay time is longer for longer wavelengths such as these. Secondly, the continuum intensity level at the end of each simulation remains elevated by about 8\% compared to $t = 0$, with all simulations converging on a similar value. Finally, it is obvious that shorter, more intense beams result in larger continuum enhancements compared to longer, milder beams. Apart from these general conclusions, we notice that case B20 behaves very similarly to case C20, meaning that an electron beam consisting of multiple triangular bursts influences the atmosphere in much the same way as a beam of constant beam flux does.\\
\indent To investigate where the excess intensity is coming from, we followed \citetalias{2026A&A...705A.157O} by making use of the contribution function to the emergent intensity. The results are shown in Fig.~\ref{fig:contrib_tri} for case T20. Similar figures for all our cases (with the exception of case C1) are included as separate panels in Fig.~\ref{fig:contrib_all}. Comparing with the \mbox{F-CHROMA} case, we notice similar trends: a substantial increase of contributions from optically thin chromospheric layers dominating during the time of beam energy input followed by a change to enhancements stemming predominantly from the photosphere during the later stages of the simulation. We notice that these photospheric enhancements are not as large as in the \mbox{F-CHROMA} case. We comment on this down below. The large increase in the contribution function at photospheric layers in the \mbox{F-CHROMA} case after about $t = 10$~s \citepalias[see Fig.~13 in][]{2026A&A...705A.157O} is also the reason that case T20 reaches its maximum earlier than the \mbox{F-CHROMA} case. Furthermore, comparing cases with a different maximum beam flux (e.g., cases C20 and C5) we see that all excess contributions (be they chromospheric or photospheric in nature) show higher amplitudes for simulations with shorter, more intense beams. This can readily be seen by comparing Fig.~\ref{fig:contrib_tri} and Fig.~\ref{fig:contrib_const_5s}. Furthermore, there is a compact feature apparent in Fig.~\ref{fig:contrib_const_5s} with an enhanced contribution to the intensity. This feature starts to appear after a few seconds, and moves downward in the atmosphere over time. We identify this as a chromospheric condensation. \citetalias{2026A&A...705A.157O} has shown such structures to make a substantial contribution to the intensity owing to their high electron densities, which is consistent with our results here. Finally, we notice another interesting aspect for the cases with constant beam flux and short beam durations (i.e., cases C10, C5, and C1): The contribution from chromospheric layers decreases noticeably below the $t = 0$ level after beam heating is turned off (see Fig.~\ref{fig:contrib_const_5s} and Fig.~\ref{fig:contrib_all}). This is a result of a decrease in the source function compared to $t = 0$.\\
\begin{figure}
    \includegraphics[width=0.46\textwidth]{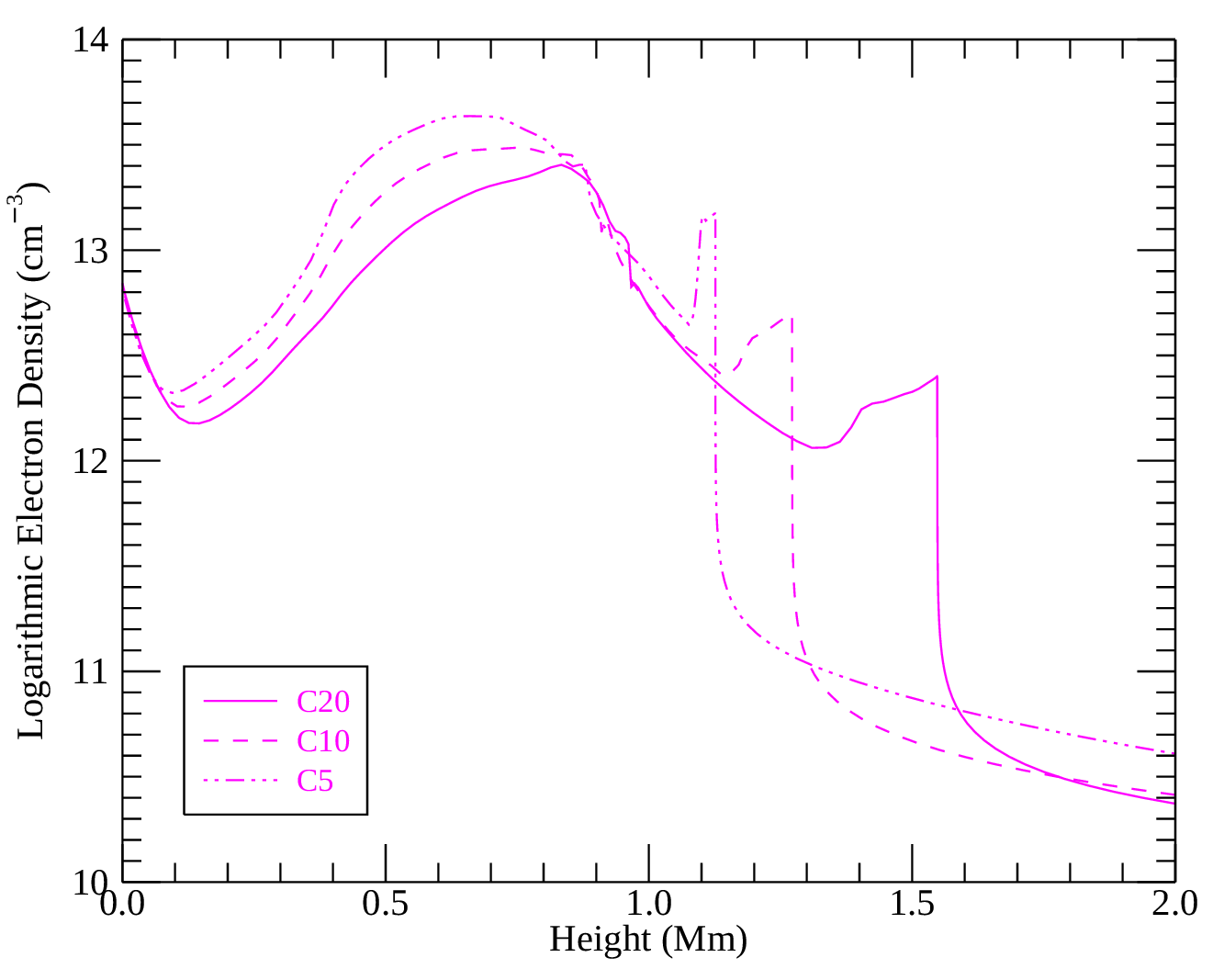}
    \caption{Logarithmic electron density as a function of height at $t = 19.9$ s for case C20 (solid line), at $t = 9.9$ s for case C10 (dashed line), and at $t = 4.9$ s for case C5 (dash-dotted line).}
    \label{fig:edens_consts}
\end{figure}
\indent The larger amplitudes of chromospheric excess contributions to the emergent intensity are related to the electron density resulting from the beam impacting on the lower atmosphere. To visualize this, we display in Fig.~\ref{fig:edens_consts} the electron density as a function of height for cases C20, C10, and C5. For each case, we take the timestep just before beam heating is turned off (which varies from case to case) in order to catch the highest electron density in each simulation. This results in a significant difference with regard to the evolution of the chromospheric evaporation front, as it has had more time to propagate in case C20 than in cases C10 and C5, leading to a substantial difference in electron density above 1.0~Mm. These regions, however, do not show obvious contribution increases, and are therefore of no interest to us. More relevant are the differences at heights between about 0.1~Mm and 0.9~Mm, where the simulations with shorter, more intense beams show higher electron densities. It follows that more electrons are available for recombination, elevating the chromospheric contribution to the excess intensity compared to cases with less intense beam heating. This result is consistent with the relationship between the electron density and the contribution to the emergent intensity found in \citetalias{2026A&A...705A.157O}, where higher electron densities were associated with larger contributions. Along the same lines, the fact that the material in and around the evaporation fronts seen in Fig.~\ref{fig:edens_consts} does not cause significant increases in the contribution to the emergent intensity (with the exception of the cores of these fronts) is also consistent with the lower limit found in \citetalias{2026A&A...705A.157O}, where \mbox{$log(n_e) \gtrsim 13$} was identified as a necessary, but not sufficient, condition for continuum enhancements to become significant. We note here that our chosen electron beam spectrum ($d = 3$ and $E_C = 25$~keV) is hard, and softer beams may not be able to reach the same depths.\\
\begin{figure}
    \includegraphics[width=0.48\textwidth]{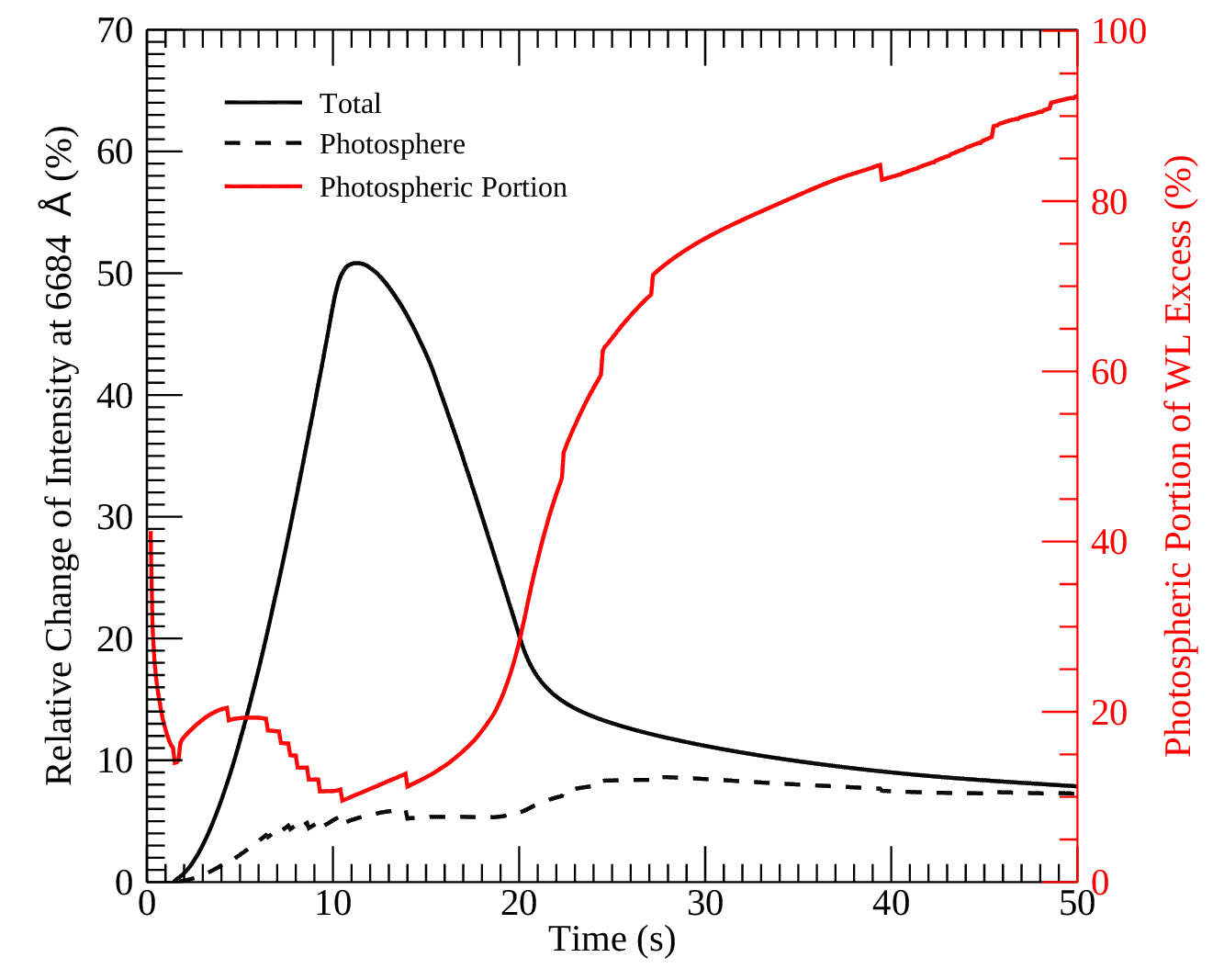}
    \caption{Relative change of the continuum intensity at 6684~\AA~and portion of excess intensity coming from the photosphere for case T20. The solid black curve depicts the light curve at 6684~\AA, while the dashed black line shows the light curve taking into account only the photospheric contribution. The percentage of the excess emissions coming from the photosphere is shown with a solid red line, and a separate y axis with the corresponding values is shown on the right.}
    \label{fig:phot_contrib}
\end{figure}
\indent We observe throughout all simulations that high continuum enhancements correlate well with enhanced chromospheric contributions. Photospheric contributions are dwarfed by the immense contributions from hydrogen recombination in the chromosphere, and only start to constitute a substantial portion of the total WL excess once the enhancements enter the declining phase (as a result of the fast decrease of excess recombination radiation). We illustrate this using case T20, as depicted in Fig.~\ref{fig:phot_contrib} (the same is shown for the other cases in Fig.~\ref{fig:phot_contrib_all}). During the time of maximum WL enhancements, the excess contribution to the intensity coming from the photosphere\footnote{We define the photosphere in this case as the regions with column masses below that of the pre-flare TMR, but which lie above the $\log \tau_{500} = 1$ layer.} is only 7.4\%. This percentage quickly increases toward the end of the declining phase of excess WL emissions, reaching 31.3\% after the decay time (see Tab.~\ref{tab:sims_wl}) and surpassing 80\% in the later stages of the simulation. Looking at the cases with a constant beam flux in Tab.~\ref{tab:sims_wl}, we notice a peculiar behavior: the shorter and more intense the energy input is, the lower the photospheric contribution to the WL excess at the time of maximum WL enhancements. For case C20, the photosphere contributes 14.4\% to the total WL excess at the maximum, for case C10 this value is 8.1\%, and for C5 it is only 4.8\%. A similar trend appears for cases T20 and T10 as well as G20$_{\mathrm{slim}}$ and G20$_{\mathrm{wide}}$. This aspect is a result of the larger chromospheric contributions for the cases with shorter, more intense beams.\\
\indent In all cases, the $\log \tau_{500} = 1$ layer has temperatures elevated by about 50~K at the end of our simulation runs. At the same time, the chromospheric contribution to the excess emission (see Fig.~\ref{fig:phot_contrib}) is low in the second half of our simulations due to the short timescale of recombination. It follows that the raised continuum level at the end of our simulations is caused by the heated photosphere. The cooling timescale of the photosphere is longer than the recombination timescale in the chromosphere, and we therefore see a long tail of slowly decaying WL excess. Similar arguments also explain the faster emission decrease in our cases compared to the \mbox{F-CHROMA} case, as the photosphere in the \mbox{F-CHROMA} simulation gets heated more strongly than in the umbral conditions under consideration here and therefore plays a more significant role during the maximum and the decay phase.\\
\begin{figure}
    {\includegraphics[width=0.48\textwidth]{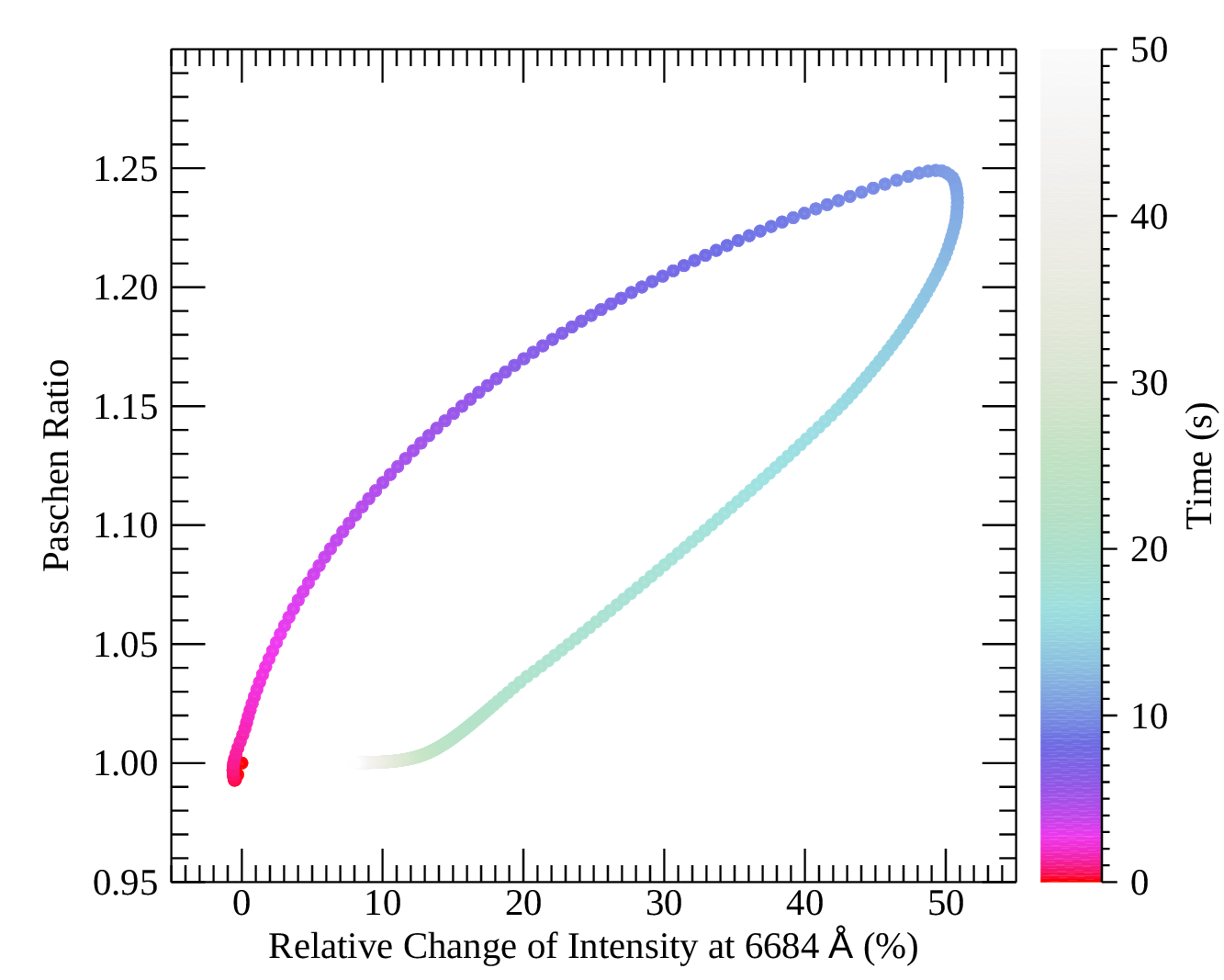}
    \includegraphics[width=0.48\textwidth]{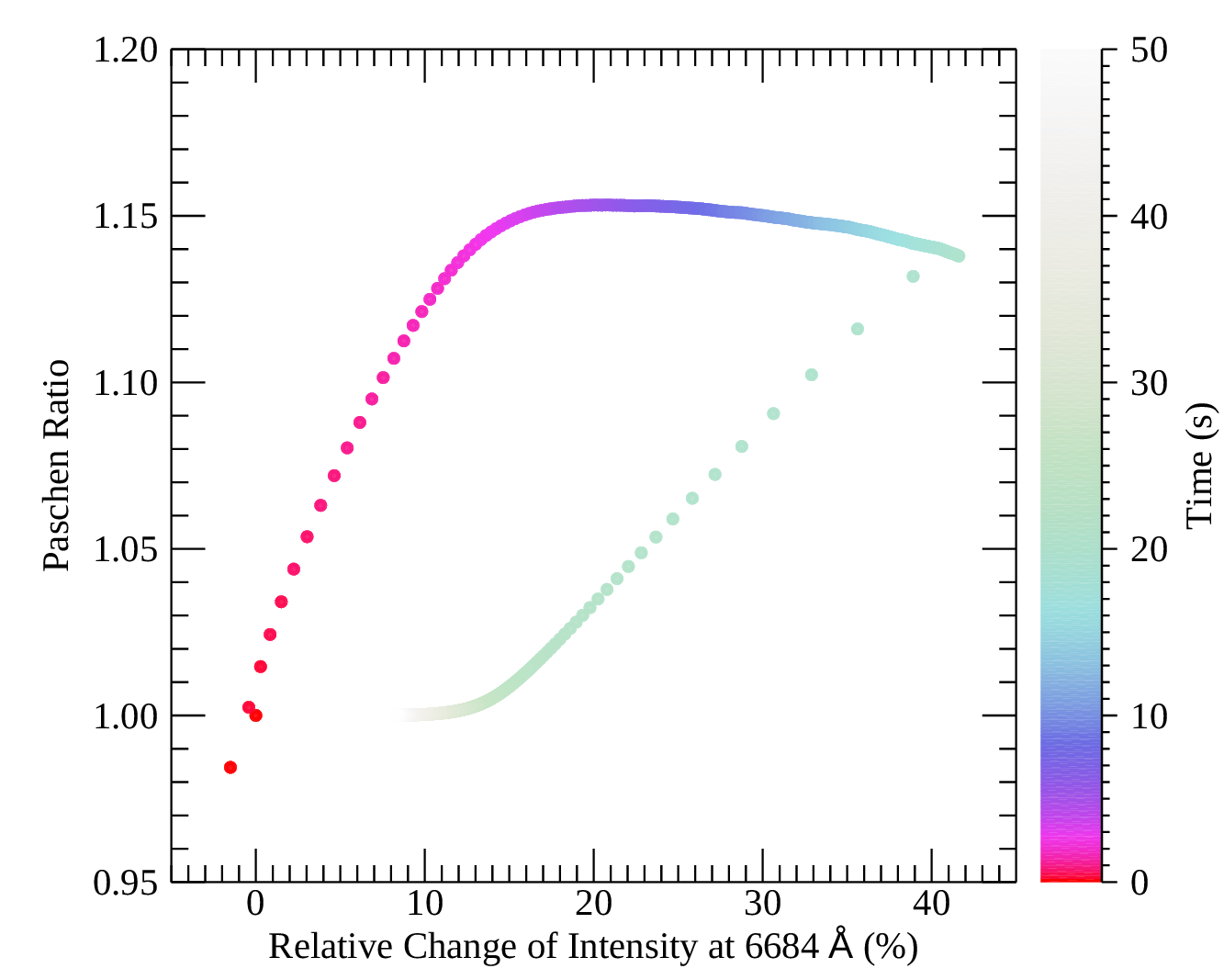}}
    \caption{Paschen ratio as a function of excess continuum intensity at 6684~\AA~for case T20 (top) and case C20 (bottom). The color of each point provides temporal information, as indicated by the color bars on the right. Note the different x axis ranges.}
    \label{fig:pratio_c6684}
\end{figure}
\renewcommand{\specialcell}[2][c]{%
\begin{tabular}[#1]{@{}c@{}}#2\end{tabular}}
\begin{table}
    \caption{Properties of the Balmer and Paschen ratio for each simulation.}\label{tab:ratios}
    \centering
    \begin{tabular}{ r r r r r }
    \hline
    Label & $\mathrm{R_{B,max}}$ & \specialcell{$\mathrm{t_{R_{B,max}}}$\\(s)} & $\mathrm{R_{P,max}}$ & \specialcell{$\mathrm{t_{R_{P,max}}}$\\(s)} \\
    \hline
    T20 & 8.05 & 10.3 & 1.25 & 10.3 \\
    T10 & 10.25 & 5.2 & 1.40 & 5.2 \\
    C20 & 5.96 & 7.4 & 1.15 & 5.5 \\
    C10 & 8.30 & 3.0 & 1.26 & 3.3 \\
    C5 & 10.68 & 1.7 & 1.42 & 2.0 \\
    C1 & 13.04 & 0.4 & 1.82 & 0.7 \\
    B20 & 6.53 & 5.8 & 1.18 & 4.8 \\
    G20$_{\mathrm{slim}}$ & 11.15 & 9.9 & 1.48 & 10.2 \\
    G20$_{\mathrm{wide}}$ & 8.14 & 10.1 & 1.25 & 10.1 \\
    \hline
    \end{tabular}
    \tablefoot{$\mathrm{R_{B,max}}$ and $\mathrm{R_{P,max}}$ describe the maximum Balmer and Paschen ratio, respectively. $\mathrm{t_{R_{B,max}}}$ and $\mathrm{t_{R_{P,max}}}$ are the corresponding times.}
\end{table}
\indent Both the Balmer and Paschen ratios are substantially higher in our cases than in the \mbox{F-CHROMA} case. The maximum values are shown in Tab.~\ref{tab:ratios}. Intuitively, the maximum ratio increases with increasing maximum beam flux. The temporal evolution compared to the continuum enhancement exhibits an elongated, loop-like shape (see Fig.~\ref{fig:pratio_c6684}, top panel) for most cases. The exceptions are the cases with constant beam fluxes: here, the ratios reach their maximum after a few seconds, followed by a slow decrease as beam heating continues. This is due to a slight decrease in the electron density in parts of the layers relevant for the enhanced hydrogen recombination radiation. The shorter, more intense the beam, the faster the ratios decrease after this maximum. In all cases, the ratios quickly approach unity after beam heating has ceased, and remain around that value during the remainder of the simulation (when the photosphere dominates the excess emission). In Fig.~\ref{fig:bratio_pratio}, we compare the Balmer and Paschen ratio directly. We chose case T10 for this due to the larger range of values covered by this simulation compared to case T20. We see that the relationship is nonlinear in nature. Similar figures for all our cases (with the exception of case C1) are included as separate panels in Fig.~\ref{fig:bratio_pratio_all}, showcasing that the range covered by each case may differ, but the qualitative behavior stays the same. The reason for the nonlinearity is the increase in the photospheric temperature combined with the original continuum intensity. The higher photospheric temperature modifies the intensity following Planck's law, affecting the continuum intensity near the Paschen jump more than near the Balmer jump. This is, however, only true for the absolute increase. The relative increase compared to the original continuum intensity (at $t = 0$) is larger near the Balmer jump, resulting in the nonlinear behavior seen in Fig~\ref{fig:bratio_pratio}. We illustrate this in Fig.~\ref{fig:intensity_change} by indicating the change in continuum intensity due to the higher photospheric temperature with two arrows. The brighter part of each arrow signifies the absolute increase in the continuum intensity, while the ratio of the bright part to the dark part marks the relative increase. The bigger relative increase around the Balmer jump then results in the nonlinear behavior seen in Fig.~\ref{fig:bratio_pratio}. Taking the intensity differences around each jump instead of the ratio (i.e., replacing the fraction in Eq.~\ref{eq:ratios} with a subtraction), meaning taking into account only the increases resulting from hydrogen recombination, the relationship is linear. Assuming that both continua form in the same region, this is a result of the different radiative rates to the first excited (for the Balmer continuum) and the second excited level (for the Paschen continuum), as shown by \cite{1990ApJ...365..391M}.
\begin{figure}
    \includegraphics[width=0.48\textwidth]{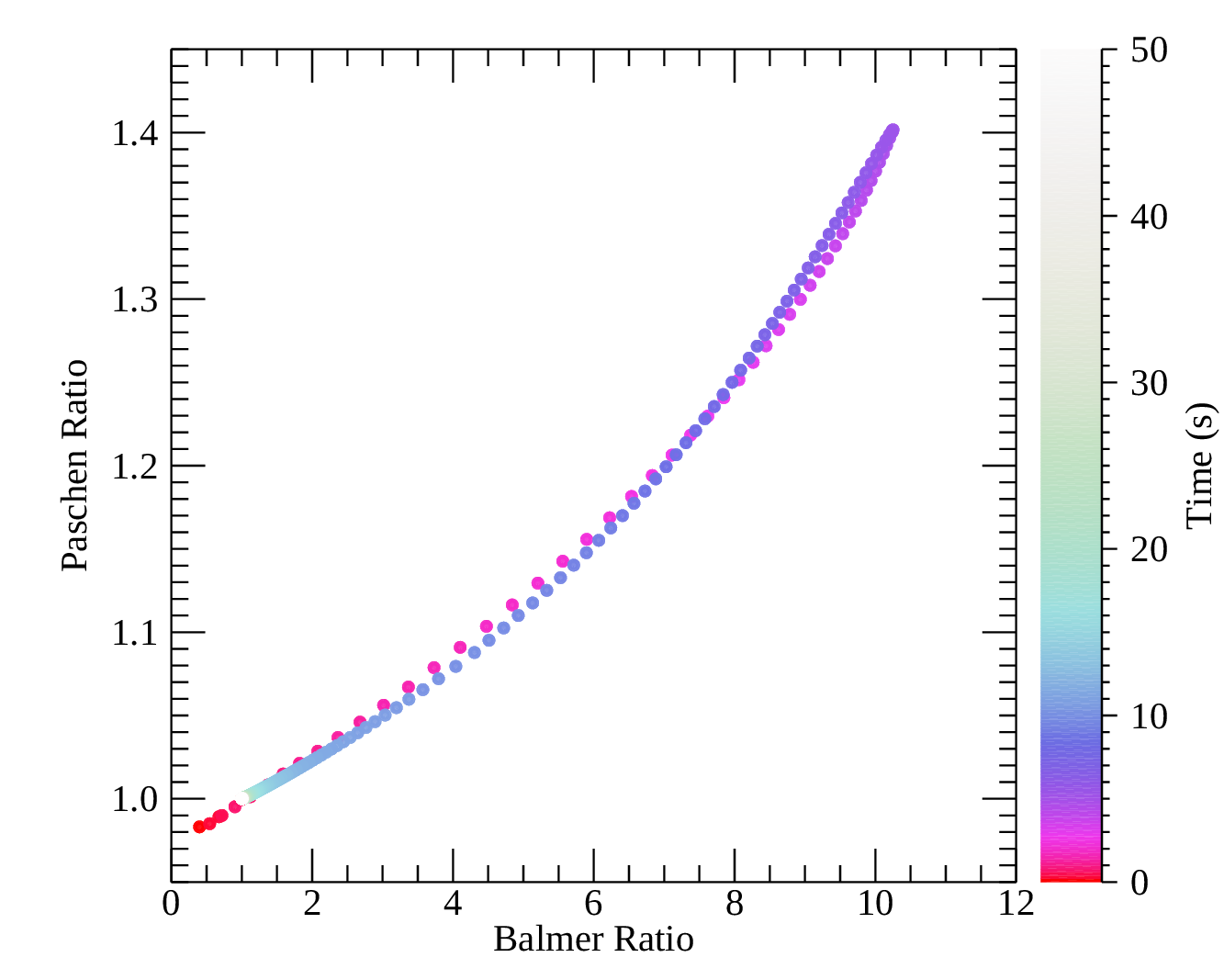}
    \caption{Paschen ratio as a function of Balmer ratio for case T10. The color of each point provides temporal information, as indicated by the color bar on the right.}
    \label{fig:bratio_pratio}
\end{figure}
\begin{figure}
    \includegraphics[width=0.48\textwidth]{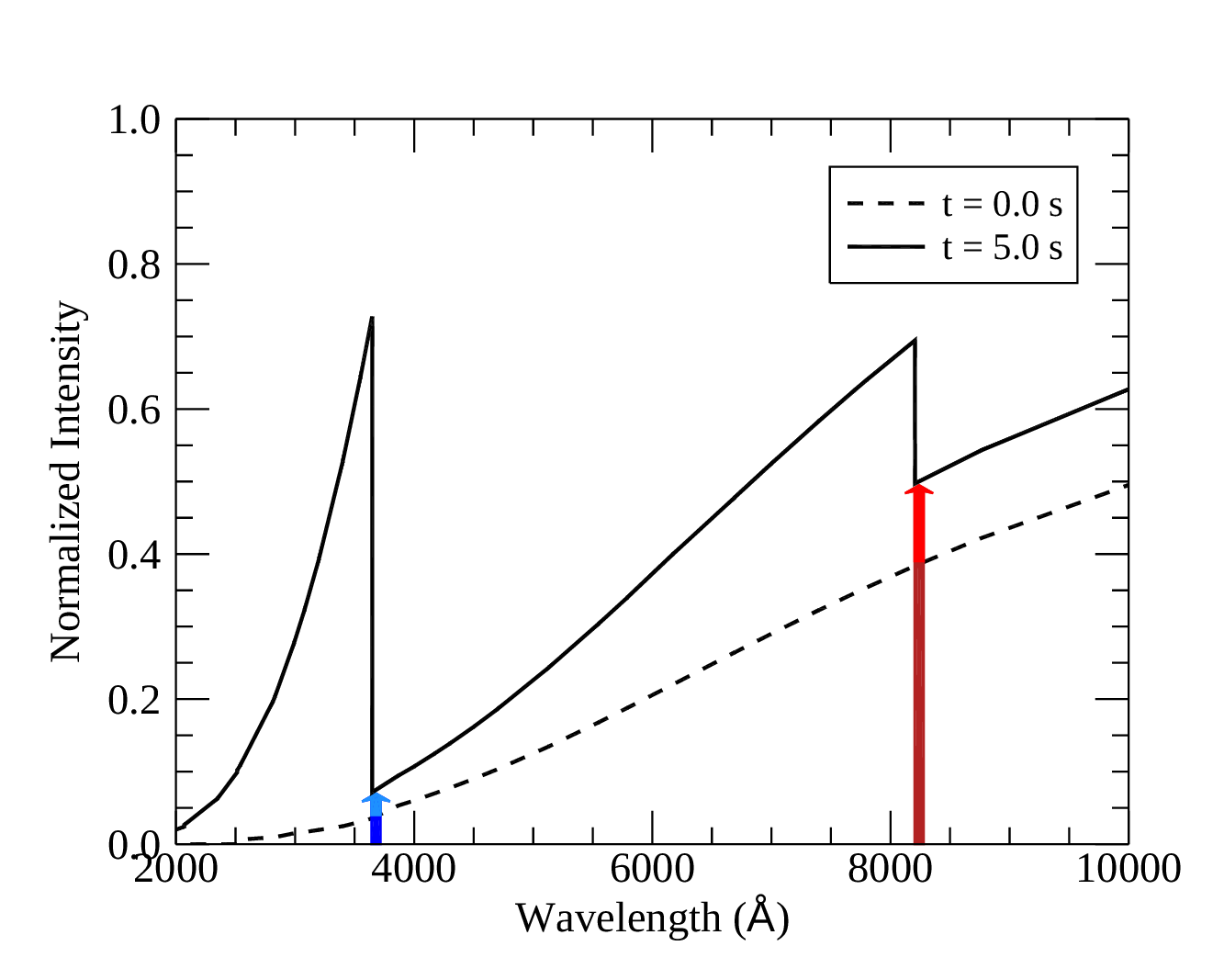}
    \caption{Normalized intensity as a function of wavelength for case T10. The dashed and solid line represent $t = 0.0$~s and \mbox{$t = 5.0$}~s, respectively. The dark blue and dark red parts of the colored arrows mark the initial continuum intensity (at $t = 0.0$~s) longward of the Balmer and Paschen jump, respectively. The bright blue and bright red parts mark the increase in the continuum intensity at $t = 5.0$~s compared to $t = 0.0$~s at either spectral point.}
    \label{fig:intensity_change}
\end{figure}


\section{Discussion}

The fact that shorter and more intense beams lead to a lower photospheric share of the total WL excess at the time of maximum continuum enhancements leads us to believe that, in order to produce type-II-like WL signatures, the electron-beam energy input in one-dimensional simulations needs to be long and gentle. Shorter, more intense beams readily ionize most of the chromospheric hydrogen, resulting in large chromospheric contributions to the excess intensity due to hydrogen recombination. On the other hand, a gentler beam may not be able to penetrate deep into the chromosphere, and it would only mildly ionize the hydrogen present there. This would, however, prevent radiative backwarming from playing a major role, presenting a challenge to our hypothesis. A test run with a constant 40-second electron beam (and a total beam energy of $10^{12}$~erg~cm$^{-2}$) does not result in the photosphere dominating the continuum enhancements at their maximum; the photospheric share is only around 25\%. Furthermore, a longer beam-energy input also delays the appearance of the maximum WL enhancements and, at least for our constant 40-second scenario here, leads to a long, gentle increase in WL emissions. It therefore remains to be seen from observations whether this scenario is plausible. We are not able to comment on whether a lower total beam energy enhances the role of the photosphere.\\
\indent The higher contribution from chromospheric layers in the simulations with more intense beams is due to a combination of two factors. First, deeper and more extensive ionization results in more ions and, crucially, more electrons available for recombination (this has already been commented on by \citealt{2017A&A...605A.125S} for infrared continuum enhancements). The electron density is higher everywhere below 1~Mm for simulations with more intense beams. Second, higher heating from the photosphere (because photospheric radiation is being stopped more efficiently due to the larger increase in chromospheric opacity) causes a redistribution of contributions from the photosphere to the chromosphere.\\
\indent Based on the analysis in this work, it may be possible to make a distinction between photospheric and chromospheric origins of WL enhancements in observations with high cadence simply by looking at the temporal behavior of the light curve: a fast decline of emissions suggests a chromospheric origin, whereas a slowly decreasing intensity level hints at substantial photospheric contributions. Observations with a cadence on the order of a few seconds may be exploitable for such analysis. We emphasize, however, that this conclusion is based on one-dimensional simulations and specific beam-energy input scenarios, and three-dimensional effects as well as different electron beams (or other energy input mechanisms) may complicate this matter for real flares.\\
\indent Even the largest intensity increases in our models do not exceed the pre-flare intensity of the models included in the \mbox{F-CHROMA} grid. We conclude that the lower level of background radiation as a consequence of the umbral atmosphere is the main reason for the large relative continuum increases, confirming the general consensus in the scientific community.


\section{Summary and conclusions}

This work investigates the influence of electron beams on an umbral atmosphere. Our results show that shorter, more intense beams lead to a faster atmospheric evolution. We identify the maximum beam flux as the dominant parameter affecting the excess WL emission: shorter, more intense beams result in larger WL enhancements. Comparing our simulation with a triangular (in time) 20-second beam to the corresponding simulation in the \mbox{F-CHROMA} grid with a similar setup (but a quiet-Sun starting atmosphere), we see a relative continuum intensity increase at 6684~\AA~that is an order of magnitude higher than in the \mbox{F-CHROMA} case. Overall, our simulations show relative continuum intensity increases at 6684~\AA~between 40 and 335\%, which is comparable to values deduced from observations. We note that the reduced umbral background intensity (compared to the quiet Sun) is the main reason for these large increases.\\
\indent Hydrogen recombination in the optically thin chromosphere is the dominant process responsible for the excess WL emissions in our cases. The photosphere gets heated via backwarming, and therefore plays the dominant role in the later stages of our simulations (once the recombination rate in the chromosphere has decreased considerably and the WL excess has decayed substantially). We note that the photospheric contribution is small in absolute intensity and strongly depends on the amplitude of chromospheric enhancements (and thus on the total energy input) driving the radiative backwarming. Shorter, more intense beams cause a larger electron density in the chromosphere, and more electrons can participate in recombination. We identify this as the reason for the larger excess WL emissions in such simulations.\\
\indent All our simulations show almost no gradual phase after beam heating is turned off. This is related to the recombination timescale in the chromosphere, combined with the fact that the chromospheric contribution to the excess intensity substantially outweighs the photospheric contribution during the time of beam heating. Since the photospheric cooling timescale is longer than the recombination timescale, the heated photosphere dominates the excess emission during the later stages of our simulations.\\
\indent In summary, we find that shorter, more intense beams are more likely to exhibit substantial WL enhancements. We conclude, however, that the underlying atmospheric conditions are of high importance for the appearance and amplitude of excess WL emissions, as well as their detectability in the first place.


\begin{acknowledgements}
    This work has been supported by the Research Council of Norway through its Centers of Excellence scheme, project number 262622. Computational resources have been provided by UNINETT Sigma2, the National Infrastructure for High-Performance Computing and Data Storage in Norway.
\end{acknowledgements}


\bibliographystyle{aa}
\bibliography{paper_bib}

\begin{appendix}

\section{Comparison of 6684~\AA~and 6175~\AA}

\begin{figure}[ht]
    \includegraphics[width=0.49\textwidth]{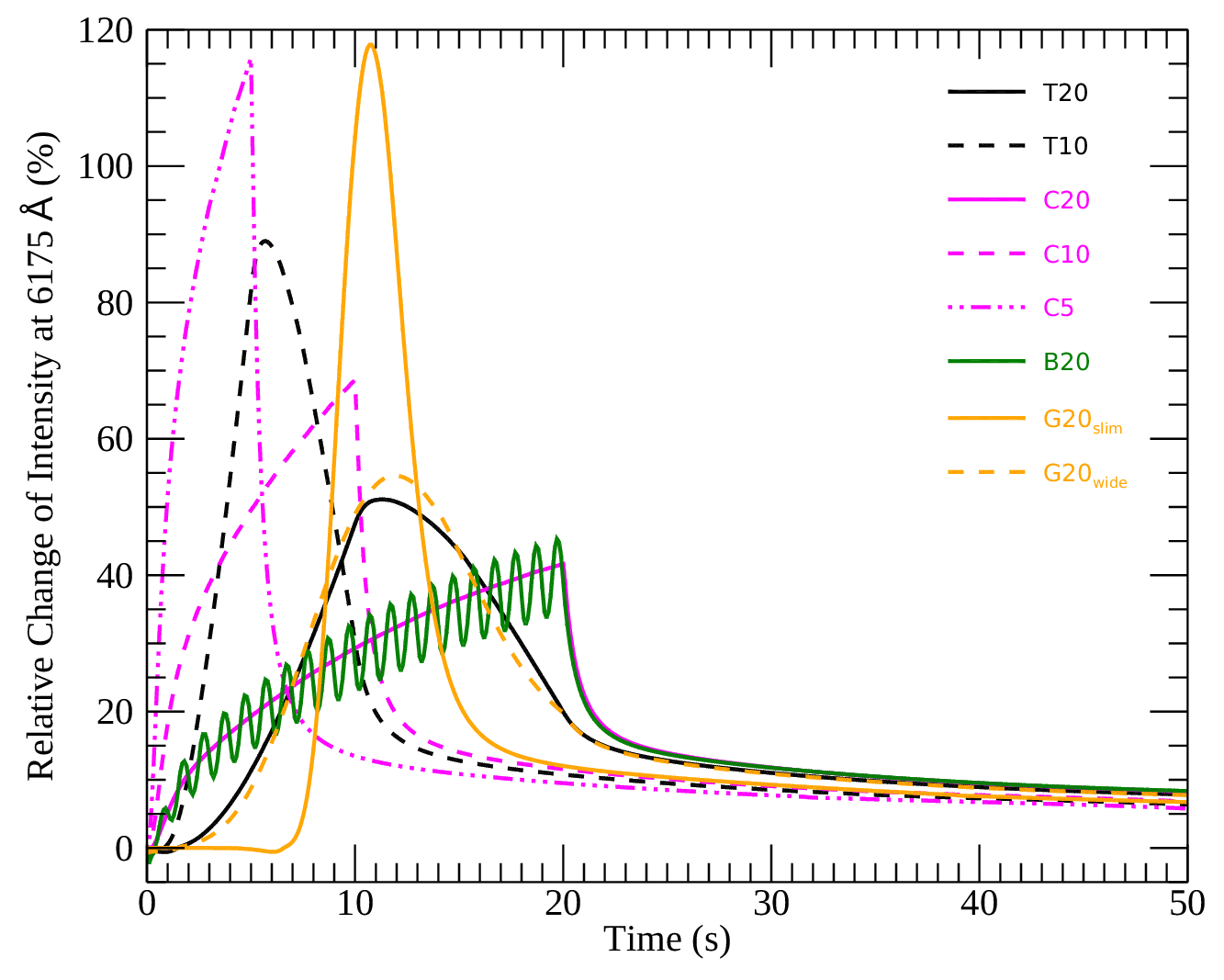}
    \caption{Continuum light curves at 6175~\AA~for all simulations in this work. Intensity increases are given with respect to the intensity at the start of the simulations. Case C1 is not included here in order to keep the y-axis range within a desirable level.}
    \label{fig:c6175}
\end{figure}

Figure \ref{fig:c6175} contains the change of the continuum intensity at 6175~\AA~as a function of time for all our simulations (excluding C1, in order to keep a reasonable y-axis range). A comparison with Fig.~\ref{fig:c6684} shows a large degree of similarity, with the respective maximum values only differing by up to two percentage points. We note  that the contribution functions behave the same at 6175~\AA~as they do for 6684~\AA~(see next section).

\section{Contribution functions}

\begin{figure}[ht]
    \includegraphics[width=0.49\textwidth]{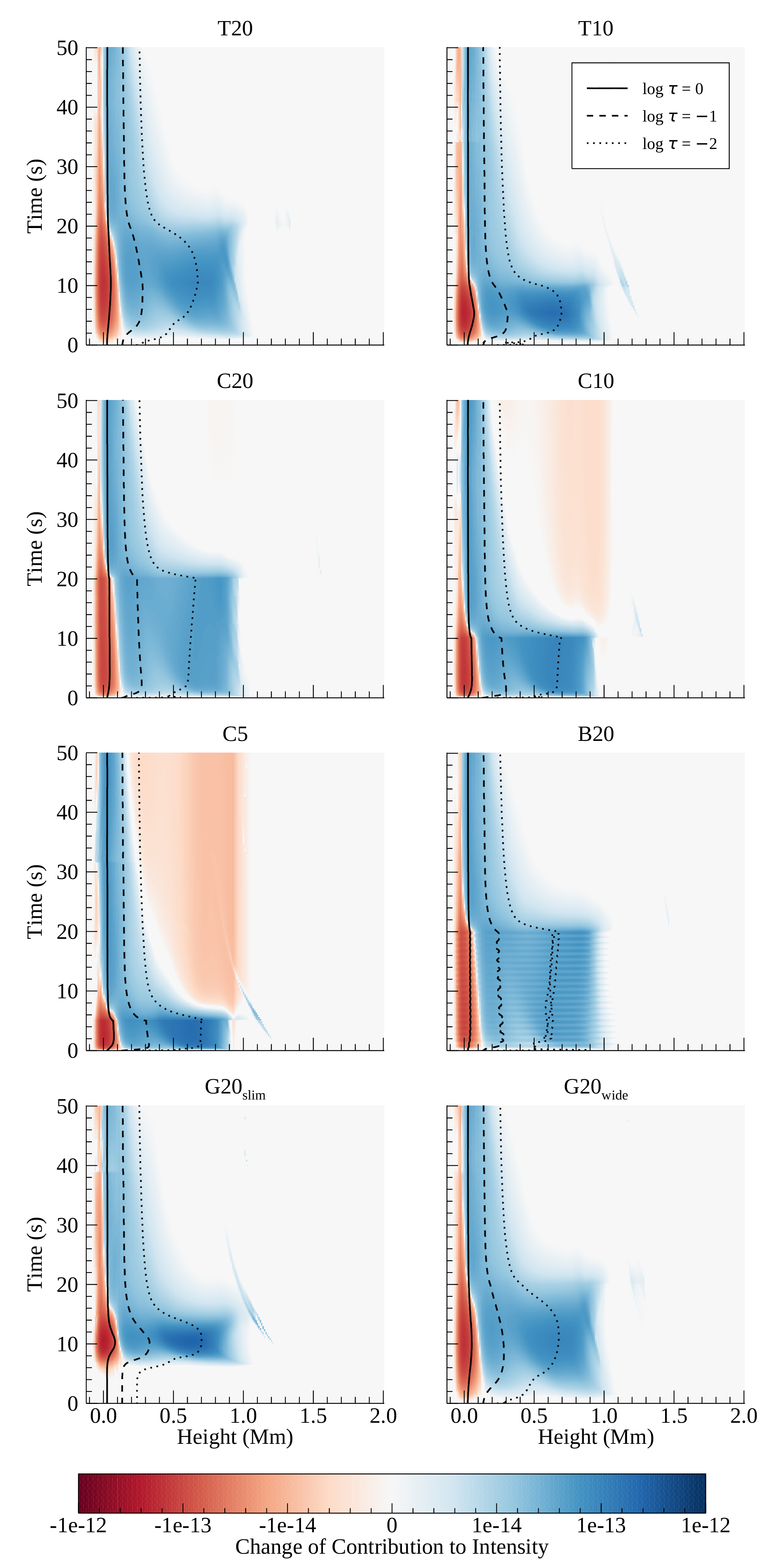}
    \caption{Change of the contribution to the continuum intensity at 6684~\AA~as a function of height and time on a logarithmic scale for all cases included in this work (excluding case C1). The unique label used to identify each simulation is given at the top of each panel. The scale is displayed as a color bar at the bottom. The bin size in the space domain is $10$~km, and $0.1$~s in the time domain. The solid, dashed, and dotted black lines in each panel indicate the height where \mbox{$\tau = 1.0$}, $0.1$, and $0.01$, respectively.}
    \label{fig:contrib_all}
\end{figure}

Figure \ref{fig:contrib_all} shows the change of the contribution function to the continuum intensity at 6684~\AA~as a function of height and time for all our simulations (excluding C1, as it has not reached the 50-second mark). All cases show a clear dominance of chromospheric excess contributions within the time of beam-energy input, with rising importance of photospheric contributions after the beam is switched off.
\clearpage

\section{Role of the photosphere in the white-light excess}

\begin{figure}[ht]
    \includegraphics[width=0.47\textwidth]{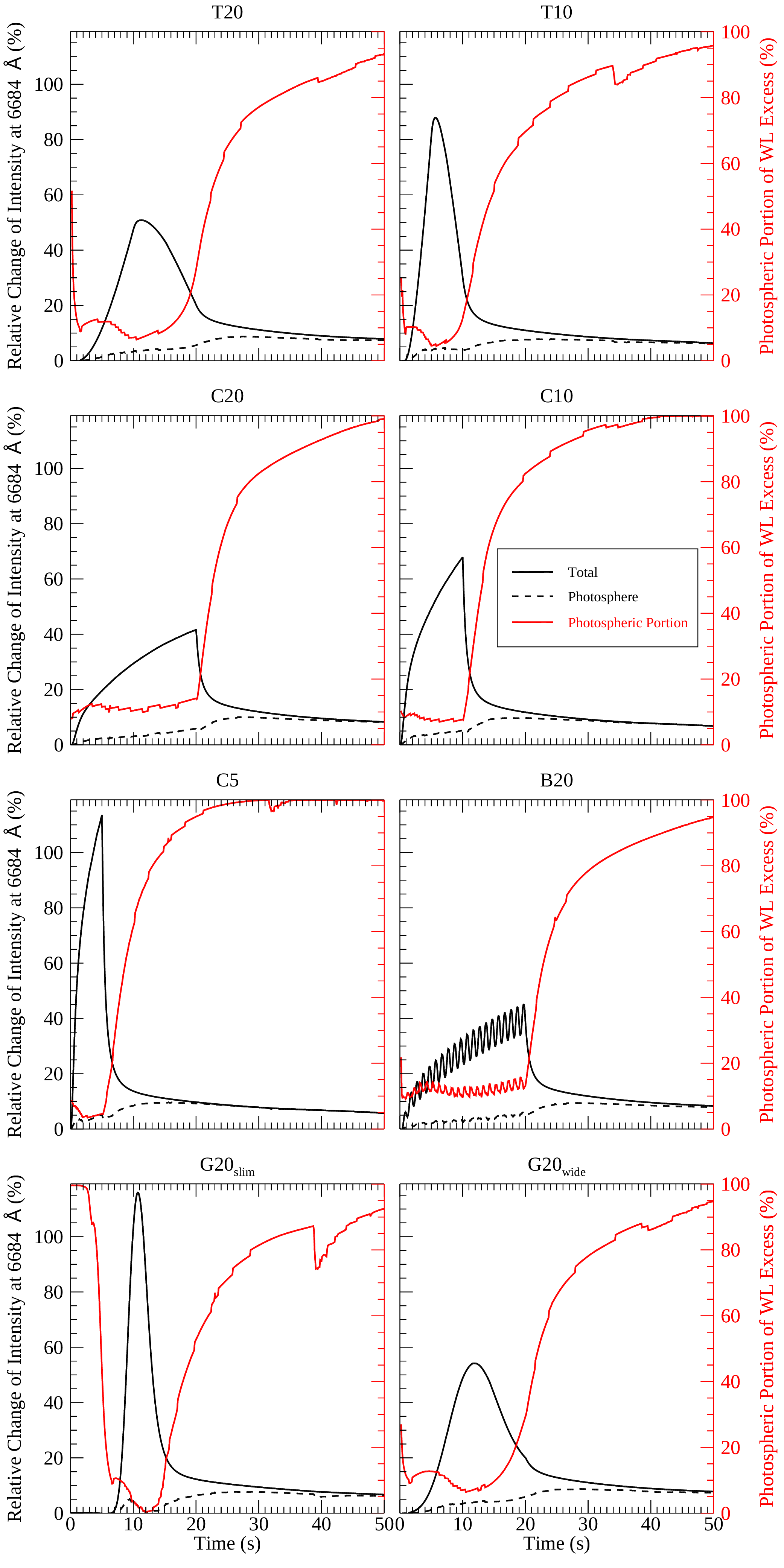}
    \caption{Relative change of the intensity at 6684~\AA~and portion of excess intensity coming from the photosphere for all cases included in this work (excluding case C1). The unique label used to identify each simulation is given at the top of each panel. The solid black curve depicts the light curve at 6684~\AA, while the dashed black line shows the light curve taking into account only the photospheric contribution. The percentage of the excess emissions coming from the photosphere is shown with a solid red line, and a separate y-axis with the corresponding values is shown on the right.}
    \label{fig:phot_contrib_all}
\end{figure}

Figure \ref{fig:phot_contrib_all} showcases the share of the excess contribution at 6684~\AA~coming from the photosphere for each case investigated in this work (excluding C1, as it has not reached the 50-second mark). The light curve at 6684~\AA~is added as a reference.

\section{Relationship between the Balmer and Paschen ratio}

\begin{figure}[ht]
    \includegraphics[width=0.49\textwidth]{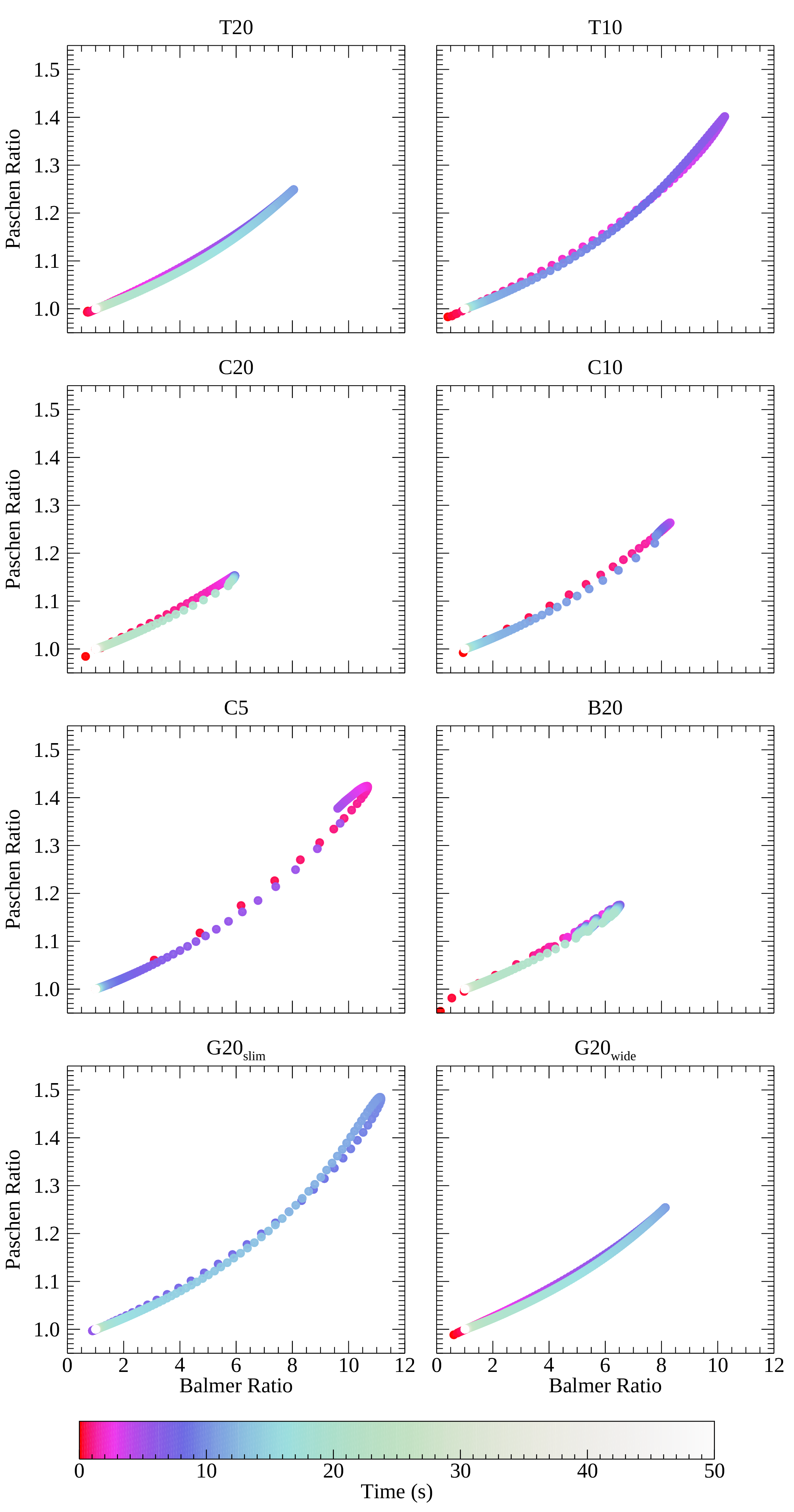}
    \caption{Paschen ratio as a function of Balmer ratio for all cases included in this work (excluding case C1). The unique label used to identify each simulation is given at the top of each panel. The color of each point provides temporal information, as indicated by the color bar at the bottom.}
    \label{fig:bratio_pratio_all}
\end{figure}

Figure \ref{fig:bratio_pratio_all} displays the relationship between the Paschen ratio and the Balmer ratio for each case investigated in this work (excluding C1, as it has not reached the 50-second mark). The relation is roughly the same for all cases and is nonlinear in nature due to the differing effect the increased photospheric temperature has on the continuum intensity as a result of Planck's law.
    
\end{appendix}

\end{document}